\documentclass[a4paper]{jpconf}
\usepackage{graphicx}
\usepackage{mathtools}
\usepackage{makecell}
\usepackage[T1]{fontenc}

\begin{document}
\title{Results of the \bfseries{\scshape{Majorana Demonstrator}}'s Search for Double-Beta Decay of $^{76}$Ge to Excited States of $^{76}$Se}


\newcommand{\itep}{National Research Center ``Kurchatov Institute'' Institute for Theoretical and Experimental Physics, Moscow, 117218 Russia}
\newcommand{\lbnl}{Nuclear Science Division, Lawrence Berkeley National Laboratory, Berkeley, CA 94720, USA}
\newcommand{\lanl}{Los Alamos National Laboratory, Los Alamos, NM 87545, USA}
\newcommand{\queens}{Department of Physics, Engineering Physics and Astronomy, Queen's University, Kingston, ON K7L 3N6, Canada}
\newcommand{\uw}{Center for Experimental Nuclear Physics and Astrophysics, and Department of Physics, University of Washington, Seattle, WA 98195, USA}
\newcommand{\unc}{Department of Physics and Astronomy, University of North Carolina, Chapel Hill, NC 27514, USA}
\newcommand{\duke}{Department of Physics, Duke University, Durham, NC 27708, USA}
\newcommand{\ncsu}{Department of Physics, North Carolina State University, Raleigh, NC 27695, USA}	
\newcommand{\ornl}{Oak Ridge National Laboratory, Oak Ridge, TN 37830, USA}
\newcommand{\ou}{Research Center for Nuclear Physics, Osaka University, Ibaraki, Osaka 567-0047, Japan}
\newcommand{\pnnl}{Pacific Northwest National Laboratory, Richland, WA 99354, USA}
\newcommand{\ttu}{Tennessee Tech University, Cookeville, TN 38505, USA}
\newcommand{\sdsmt}{South Dakota School of Mines and Technology, Rapid City, SD 57701, USA}
\newcommand{\usc}{Department of Physics and Astronomy, University of South Carolina, Columbia, SC 29208, USA}
\newcommand{\usd}{Department of Physics, University of South Dakota, Vermillion, SD 57069, USA}  
\newcommand{\ut}{Department of Physics and Astronomy, University of Tennessee, Knoxville, TN 37916, USA}
\newcommand{\tunl}{Triangle Universities Nuclear Laboratory, Durham, NC 27708, USA}
\newcommand{\mpi}{Max-Planck-Institut f\"{u}r Physik, M\"{u}nchen, 80805 Germany}
\newcommand{\tum}{Physik Department and Excellence Cluster Universe, Technische Universit\"{a}t, M\"{u}nchen, 85748 Germany}
\newcommand{\MIT}{Department of Physics, Massachusetts Institute of Technology, Cambridge, MA 02139, USA} 
\newcommand{\williams}{Department of Physics, Williams College, Williamstown, MA 01267, USA}

\author{I.S.~Guinn$^{1,18}$, I.J.~Arnquist$^2$, F.T.~Avignone~III$^{10,15}$, A.S.~Barabash$^3$, C.J.~Barton$^{16}$, F.E.~Bertrand$^{10}$, B.~Bos$^{1,14,18}$, M.~Busch$^{7,18}$, M.~Buuck$^8$, T.S.~Caldwell$^{1,18}$, Y-D.~Chan$^4$, C.D.~Christofferson$^{14}$, P-H.~Chu$^5$, M.L.~Clark$^{1,18}$, C.~Cuesta$^8$, J.A.~Detwiler$^8$, A.~Drobizhev$^4$, D.W.~Edwins$^{15}$, Yu.~Efremenko$^{10,17}$, H.~Ejiri$^{11}$, S.R.~Elliott$^5$, T.~Gilliss$^{1,18}$, G.K.~Giovanetti$^{22}$, M.P.~Green$^{9,10,18}$, J.~Gruszko$^{21}$, V.E.~Guiseppe$^{10}$, C.R.~Haufe$^{1,18}$, R.J.~Hegedus$^{1,18}$, R.~Henning$^{1,18}$, D.~Hervas~Aguilar$^{1,18}$, E.W.~Hoppe$^2$, A.~Hostiuc$^8$, M.F.~Kidd$^{13}$, I.~Kim$^5$, R.T.~Kouzes$^2$, A.M.~Lopez$^{17}$, J.M. L\'opez-Casta\~no$^{16}$, E.L.~Martin$^{1,18}$, R.D.~Martin$^6$, R.~Massarczyk$^5$, S.J.~Meijer$^5$, S.~Mertens$^{19,20}$, J.~Myslik$^4$, T.K.~Oli$^{16}$, G.~Othman$^{1,18}$, W.~Pettus$^8$, A.W.P.~Poon$^4$, D.C.~Radford$^{10}$, J.~Rager$^{1,18}$, A.L.~Reine$^{1,18}$, K.~Rielage$^5$, N.W.~Ruof$^8$, M.J.~Stortini$^5$, D.~Tedeschi$^{15}$, R.L.~Varner$^{10}$, J.F.~Wilkerson$^{1,10,18}$, C.~Wiseman$^8$, W.~Xu$^{16}$, C.-H.~Yu$^{10}$, B.X.~Zhu$^5$}
\address{$^1$\unc}
\address{$^2$\pnnl}
\address{$^3$\itep}
\address{$^4$\lbnl}
\address{$^5$\lanl}
\address{$^6$\queens}
\address{$^7$\duke}
\address{$^8$\uw}
\address{$^9$\ncsu}
\address{$^{10}$\ornl}
\address{$^{11}$\ou}
\address{$^{12}$\pnnl}
\address{$^{13}$\ttu}
\address{$^{14}$\sdsmt}
\address{$^{15}$\usc}
\address{$^{16}$\usd}
\address{$^{17}$\ut}
\address{$^{18}$\tunl}
\address{$^{19}$\mpi}
\address{$^{20}$\tum}
\address{$^{21}$\MIT}
\address{$^{22}$\williams}


\ead{iguinn@email.unc.edu}

\begin{abstract}
  The \textsc{Majorana Demonstrator} is searching for double-beta decay of $^{76}$Ge to excited states (E.S.) in $^{76}$Se using a modular array of high purity Germanium detectors. $^{76}$Ge can decay into three E.S.s of $^{76}$Se. The E.S. decays have a clear event signature consisting of a $\beta\beta$-decay with the prompt emission of one or two $\gamma$-rays, resulting in with high probability in a multi-site event. The granularity of the \textsc{Demonstrator} detector array enables powerful discrimination of this event signature from backgrounds. Using 21.3~kg-y of isotopic exposure, the \textsc{Demonstrator} has set world leading limits for each E.S. decay, with 90\% CL lower half-life limits in the range of $(0.56-2.1)\cdot10^{24}$~y. In particular, for the $2\nu$ transition to the first $0^+$ E.S. of $^{76}$Se, a lower half-life limit of $0.68\cdot10^{24}$ at 90\% CL was achieved.  
\end{abstract}

\section{Introduction}
\begin{figure}[b]
  \centering
  \includegraphics[width=0.5\textwidth]{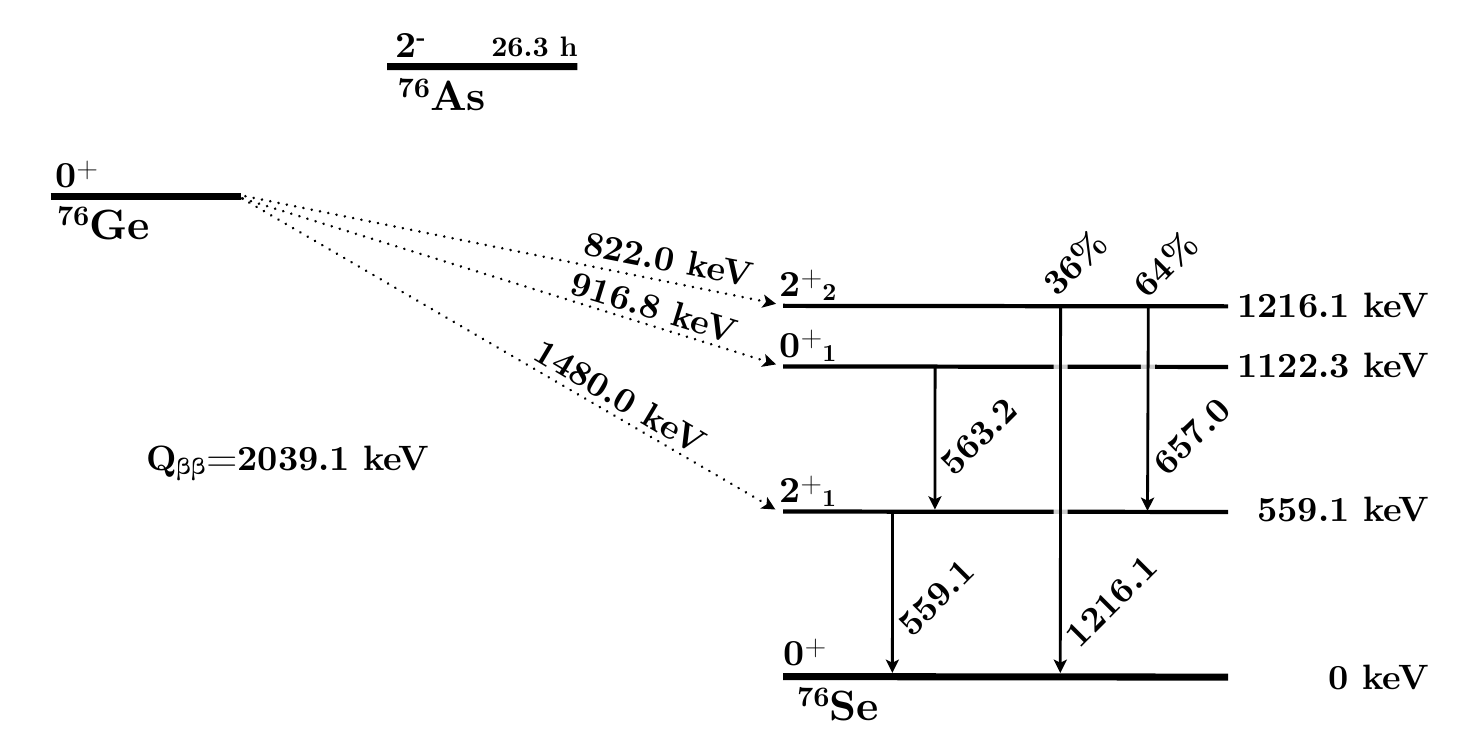}
  \caption{\label{leveldiagram}Level diagram of the $\beta\beta$-decay of $^{76}$Ge into $^{76}$Se.}
\end{figure}
Double-beta ($\beta\beta$) decay is the subject of a varied experimental program.
Of particular interest is neutrinoless double-beta decay ($0\nu\beta\beta$), a hypothetical lepton number violating process that would indicate that the neutrino is a Majorana fermion.
While two-neutrino double-beta ($2\nu\beta\beta$) decay has been directly observed in eleven isotopes, $0\nu\beta\beta$ has not yet been observed.
Most $\beta\beta$ isotopes have multiple daughter excited states (E.S.) that they can decay into; for example, the decay scheme of $^{76}$Ge is shown in Figure~\ref{leveldiagram}.
Such decays will have a reduced Q-value, with the remaining decay energy emitted as one or more $\gamma$s.
These $\gamma$s will typically travel a few centimeters before interacting in bulk material, resulting in multiple energy deposition sites.

Measurement of the half-life of $2\nu\beta\beta$ to E.S. provides a useful check on the accuracy of $2\nu\beta\beta$ nuclear matrix element calculations; in turn, this may help evaluate the $0\nu\beta\beta$ matrix elements.
In addition, $\beta\beta$ decays to E.S. are sensitive to physics beyond the standard model.
The half-life of $0\nu\beta\beta$ decay to E.S. is sensitive to the underlying physics mechanism \cite{simkovic}.
The half-life of $2\nu\beta\beta$ decay to $2^+$ E.S. is sensitive to a hypothetical bosonic component to neutrinos \cite{dolgov, barabash1}.
So far, $2\nu\beta\beta$ to the first $0^+$ daughter E.S. has been observed in $^{100}$Mo and $^{150}$Nd, with a tension between these observations and theoretical half-life estimates \cite{barabash2}.

\section{The MAJORANA DEMONSTRATOR}
\begin{figure}
  \centering
  \begin{minipage}{0.5\textwidth}
    \centering
    \includegraphics[height=3.1cm]{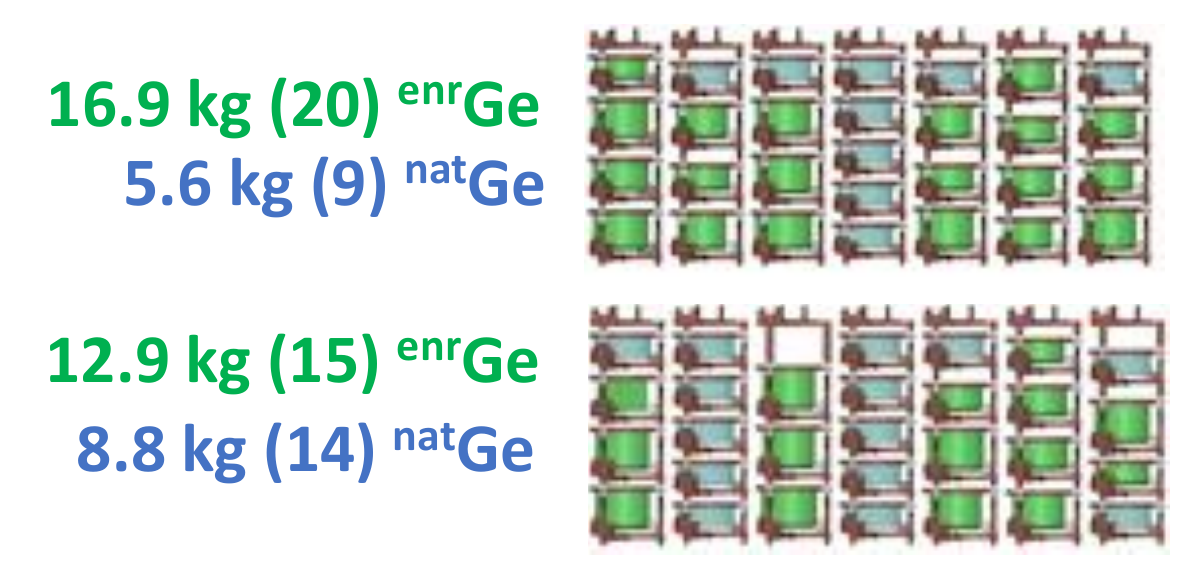}
    \caption{\label{modules}A drawing of the content of module~1~(top) and~2~(bottom). Blue detectors have the natural $^{76}$Ge isotopic abundance, while green are enriched to 88\%.}
  \end{minipage}\hspace{2pc}%
  \begin{minipage}{0.4\textwidth}
    \centering
    \includegraphics[height=3.5cm]{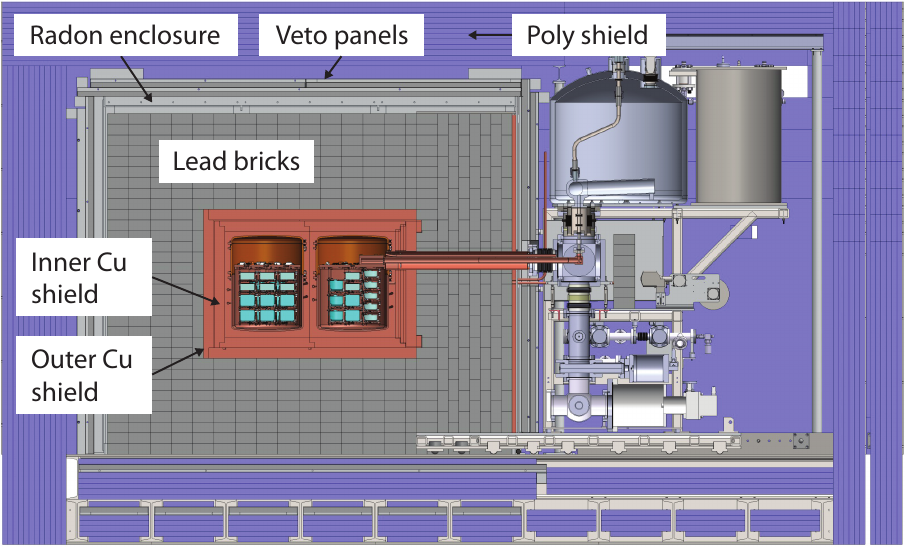}
    \caption{\label{mjd}A drawing of the \textsc{Majorana Demonstrator} shield, with modules inserted.}
  \end{minipage} 
\end{figure}

The \textsc{Majorana Demonstrator} \cite{mjd} is searching for $0\nu\beta\beta$ of $^{76}$Ge using HPGe detectors.
The experiment consists of two separate modules, consisting of arrays of detectors operated in separate cryostats in vacuum.
44.1~kg of detectors are used, 29.7~kg of which are enriched to 88\% in $^{76}$Ge, allowing them to act as both the source and dector of $0\nu\beta\beta$.
The detectors in each module are shown in Figure~\ref{modules}.
In order to minimize backgrounds, the experiment is constructed using ultra-low background materials and is placed within a multi-layered shield shown in Figure~\ref{mjd}.
The experiment is housed at the 4850' level (4300 m.w.e) of the Sanford Underground Research Facility (SURF) in order to minimize exposure to cosmic ray muons; in addition, scintillating polyethylene veto panels surround the experiment and actively veto additional muons.
The HPGe detectors use the P-type Point Contact (PPC) detector geometry, which has advantages in energy resolution.
The PPC geometry and the granularity of the detector array enable discrimination of single- and multi-site events.

The \textsc{Majorana Demonstrator} is also searching for $\beta\beta$ decay to excited states in $^{76}$Ge, which has not yet been observed in this isotope.
$^{76}$Se has three E.S. that $^{76}$Ge can decay into, as shown in Figure~\ref{leveldiagram}.
$2\nu\beta\beta$ to the $0^+_1$ state has the shortest expected half-life, with theoretical estimates ranging from $1.0\cdot10^{23}-7.1\cdot10^{24}$~y \cite{barabash2}; since the \textsc{Demonstrator} has a sensitivity within this range of half-lives, this state will be the focus of this document.
The $0^+_1$ E.S. decay mode has a Q-value of 917~keV and two $\gamma$s, with energies 559~keV and 563~keV.
Because $\beta\beta$ to E.S. is an inherently multi-site event, the \textsc{Demonstrator} can significantly reduce its backgrounds by searching only for events that involve multiple detector hits.
In particular, we will search for peaks at 559~keV and 563~keV in individual detectors within multi-detector events.
Multi-detector events are built to include all detector hits within a 4~$\mu$s rolling window; this is a conservative window that is expected to capture all truly simultaneous detector hits.

\section{Background Cuts and Detection Efficiency}
This analysis uses the standard channel selection, data cleaning, and muon cuts developed for the \textsc{Majorana Demonstrator}'s $0\nu\beta\beta$ analysis \cite{mjd}.
In addition, the detector hits in coincidence with candidate gamma hits will provide additional observables that can be used to further cut backgrounds.
Since $95\%$ of $^{76}$Ge is contained in enriched detectors and the $\beta\beta$ site will be contained in a separate detector from the gamma peak, events in which none of the coincident detectors are enriched are cut.

\begin{figure}
  \centering
  \includegraphics[width=0.45\textwidth]{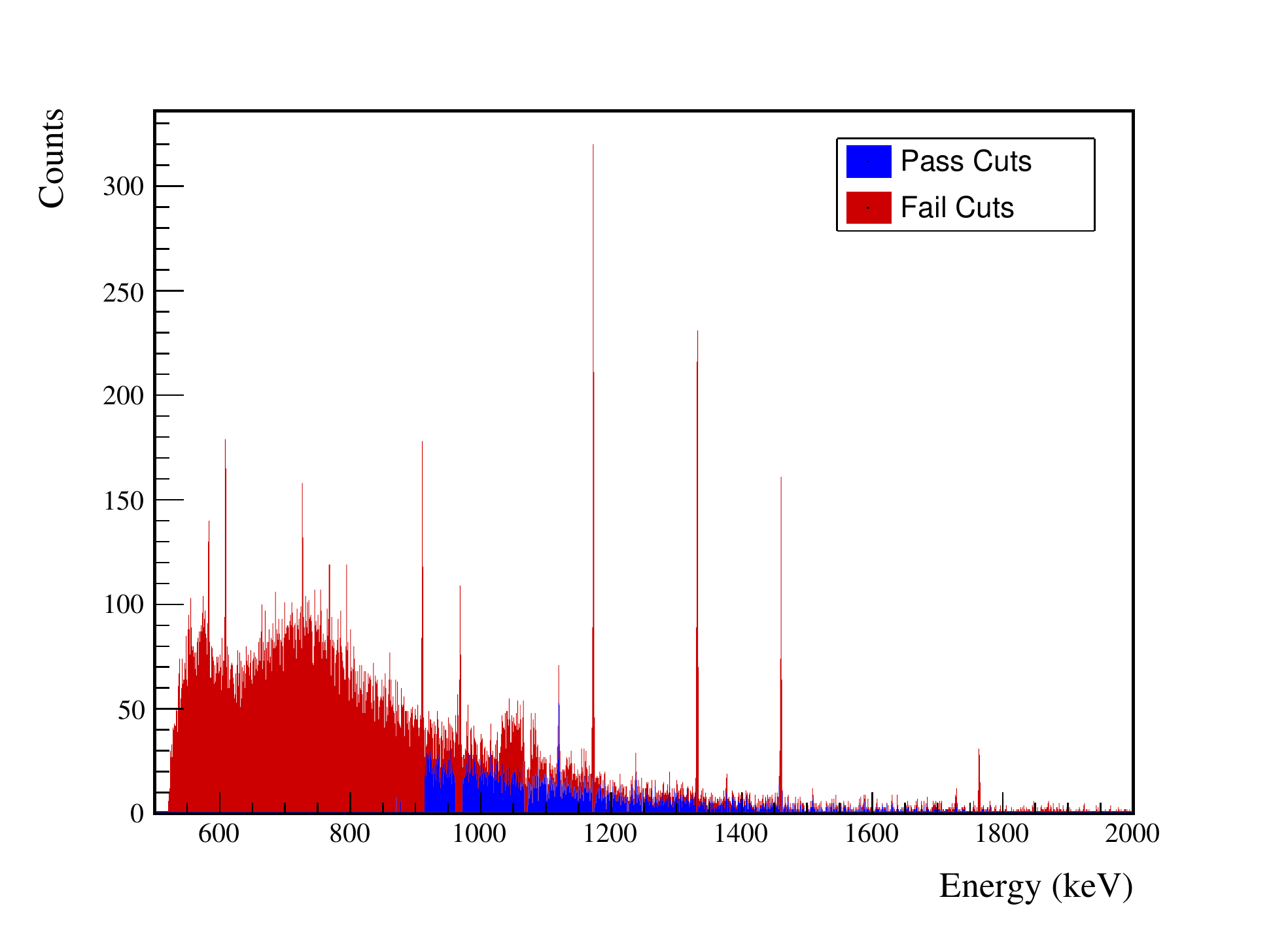}
  \includegraphics[width=0.45\textwidth]{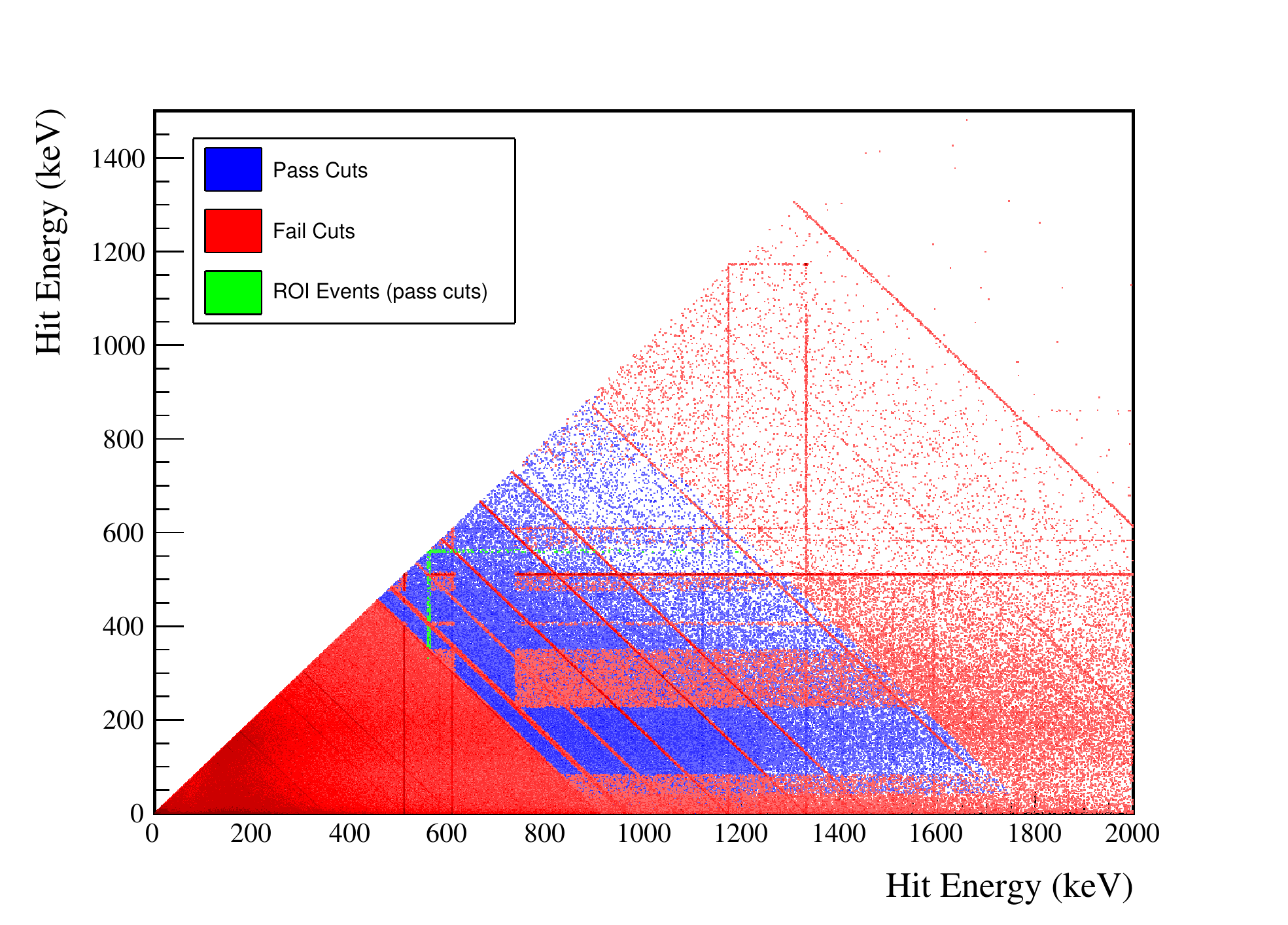}
  \caption{\label{bgcuts}Effect of sum and coincident energy cuts on a simulated background model. Left: Sum energy spectrum of multi-detector events. Right: energy spectrum of multiplicity 2 events.}
\end{figure}

The largest source of backgrounds for this analysis is $\gamma$ rays, which are often multi-site events.
Because background $\gamma$s will mostly originate from a handful of known isotopes, they can be cut based on the energies of coincident detector hits.
Multi-detector events caused by $\gamma$ cascades will commonly produce coincidence hits with known energies.
Compton scattered $\gamma$s will commonly produce multi-detector events where the sum of energies between detectors has a known energy.
For this reason, the sensitivity of the analysis can be improved by selecting a set of energy ranges for coincident detectors and for the sum of all detectors to cut.
The energy windows are algorithmically selected in order to optimize the detection sensitivity based on simulations of the backgrounds and each E.S. decay mode.
Figure~\ref{bgcuts} shows the energy ranges selected for the first $0^+$ $^{76}$Se E.S.
The expected effect of these energy cuts is to improve the signal to background ratio by a factor of 3.9 for the 559~and 563~keV $\gamma$ peaks.

MAGE \cite{mage}, a GEANT4 \cite{geant} based simulation package containing a simulated geometry of the \textsc{Majorana Demonstrator}, combined with the DECAY0 \cite{decay0} $\beta\beta$-decay event generator, were used to produce simulations of each E.S. decay mode for $^{76}$Ge.
Using these simulations, the detection efficiency for the first $0^+$ $^{76}$Se E.S. decay mode was estimated to be $2.3\pm0.2$\% for module~1 and $1.0\pm0.2$\% for module~2.

To test the accuracy of the simulated detection efficiencies, pair production events produced by a $^{56}$Co calibration source acted as a proxy for the $\beta\beta$ E.S. events.
Multi-detector events with a 511~keV annihilation $\gamma$ detected in coincidence with single- and double-escape peak (SEP and DEP) events are used to immitate the detection signature; single-detector events in the SEP and DEP are used to control for the total rate of these events in order to estimate a detection efficiency.
The detection efficiency is measured from $^{56}$Co calibration data and simulations in 16 different SEP and DEPs.
The differences are used to estimate the systematic error associated with the simulated detection efficiencies.
The dominant systematic uncertainties originate from the detector dead layer thickness and inaccuracies in the simulation.

\section{Results}
\begin{figure}
  \centering
  \includegraphics[width=0.45\textwidth]{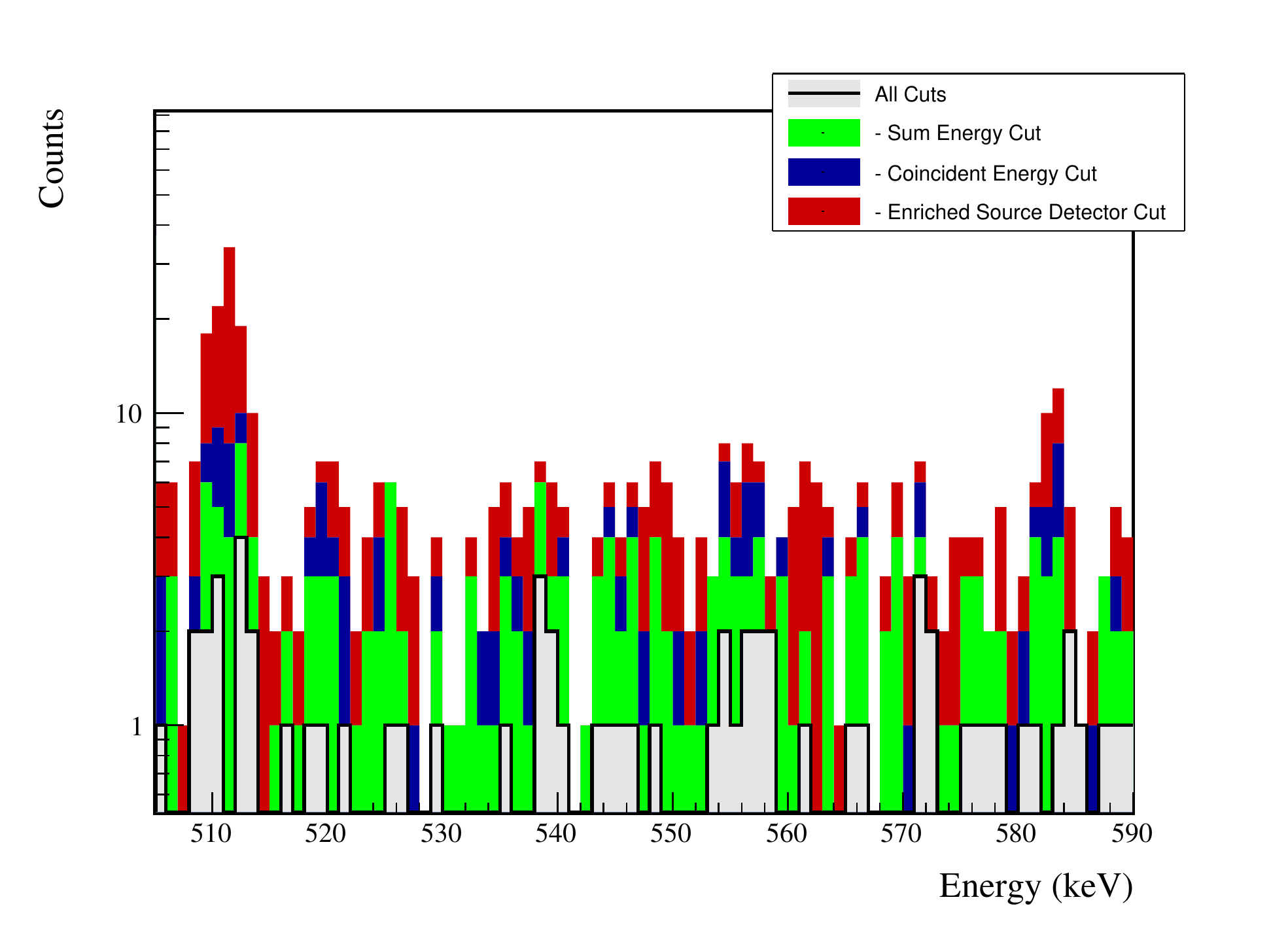}
  \includegraphics[width=0.45\textwidth]{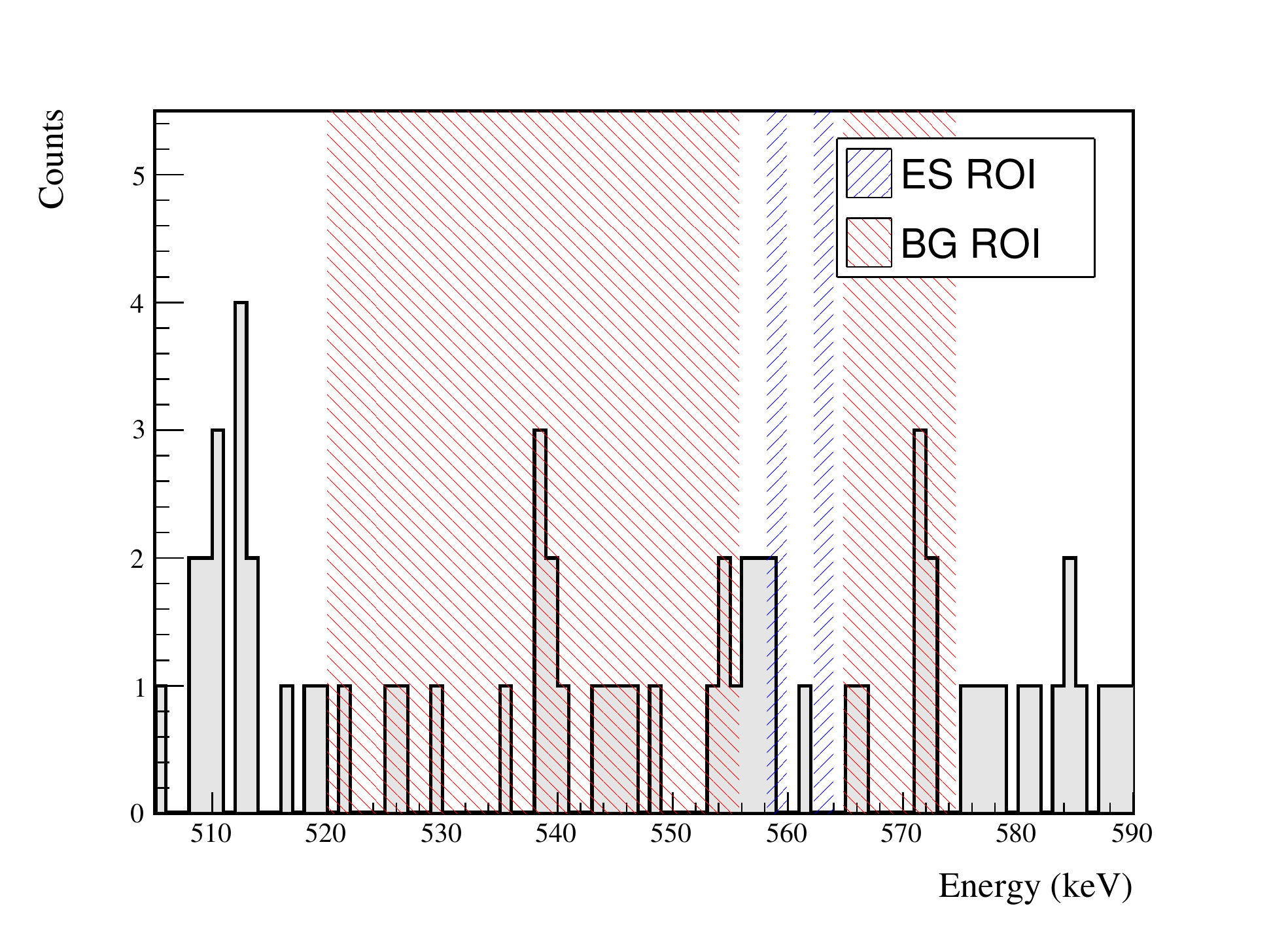}
  \caption{\label{dataspectrum}Measured background spectrum with each cut sequentially applied (left), and background and signal ROIs labelled (right).}
\end{figure}
A 90\% confidence limit was set for each E.S. decay mode using frequentist techniques.
Open data recorded in both modules between January 12, 2016 and April 18, 2018 were used for this analysis, with a total isotopic exposure of 21.3~kg-y.
A Poisson peak counting analysis was performed; for the first $0^+$ E.S., the combined regions of interest, consisting of a 1.6~keV range around the 559~and 563~keV $\gamma$ peaks, 2~counts were measured, with 2.02~background counts expected.
Figure~\ref{dataspectrum} shows multi-site hits after all cuts around these ROIs.
For each decay mode, a world leading limit was set, as shown in Table~\ref{limits}.
\begin{table}[h]
  \centering
  \caption{\label{limits} Table of limits at 90\% CL for each $\beta\beta$ to E.S. decay mode}
  \begin{tabular}{|c|c|c|c|c|}
    \hline
    Decay Mode & Q-value & $\gamma$ Energies & Previous Limit & MJD Limit \\
    \hline
    $0^+_{g.s.} \xrightarrow{2\nu\beta\beta} 0^+_1$ & 916.8 keV & 559.1+563.2 keV & $3.7\cdot10^{23}$ y \cite{gerda} & $6.8\cdot10^{23}$ y \\
    $0^+_{g.s.} \xrightarrow{2\nu\beta\beta} 2^+_1$ & 1480.0 keV & 559.1 keV & $1.6\cdot10^{23}$ y \cite{gerda} & $9.6\cdot10^{23}$ y \\
    $0^+_{g.s.} \xrightarrow{2\nu\beta\beta} 2^+_2$ & 822.0 keV & \makecell{559.1+657.0 keV\\or 1216.1 keV} & $2.3\cdot10^{23}$ y \cite{gerda} & $5.6\cdot10^{23}$ y \\
    $0^+_{g.s.} \xrightarrow{0\nu\beta\beta} 0^+_1$ & 916.8 keV & 559.1+563.2 keV & $1.3\cdot10^{22}$ y \cite{morales} & $2.1\cdot10^{24}$ y \\
    $0^+_{g.s.} \xrightarrow{0\nu\beta\beta} 2^+_1$ & 1480.0 keV & 559.1 keV & $1.3\cdot10^{23}$ y \cite{maier} & $1.1\cdot10^{24}$ y \\
    $0^+_{g.s.} \xrightarrow{0\nu\beta\beta} 2^+_2$ & 822.0 keV & \makecell{559.1+657.0 keV\\or 1216.1 keV} & $1.4\cdot10^{21}$ y \cite{barabash3} & $1.6\cdot10^{24}$ y \\
    \hline
  \end{tabular}
\end{table}
  
This material is based upon work supported by the U.S. Department of Energy, Office of Science, Office of Nuclear Physics, the Particle Astrophysics and Nuclear Physics Programs of the National Science Foundation, the Russian Foundation for Basic Research, the Natural Sciences and Engineering Research Council of Canada, the Canada Foundation for Innovation John R.~Evans Leaders Fund, the National Energy Research Scientific Computing Center, and the Oak Ridge Leadership Computing Facility, and the Sanford Underground Research Facility.

\end{document}